\documentclass[aps,reprint,prl,superscriptaddress,nofootinbib,preprintnumbers]{revtex4-2}

\usepackage[utf8]{inputenc}
\usepackage{mathtools}
\usepackage{amsmath}
\usepackage{amsthm}
\usepackage{amssymb}
\usepackage{setspace}
\usepackage{microtype}
\usepackage{antmath}
\usepackage{stmaryrd}
\usepackage{slashed}
\usepackage{xcolor}
\definecolor{linkcolor}{rgb}{0,0,0.4}
\usepackage[pdftex,colorlinks=true,
pdfstartview=FitV,
linkcolor= linkcolor,
citecolor= linkcolor,
urlcolor= linkcolor,
hyperindex=true,
hyperfigures=false]
{hyperref}

\makeatletter
\usepackage{tikz}
\usepackage{tikz-cd}
\usetikzlibrary{backgrounds}
\usepackage[framemethod=tikz]{mdframed}
\AtBeginEnvironment{mdframed}{%
\tikzset{every picture/.style={}}%
}
\mdfsetup{roundcorner=.5ex}

\theoremstyle{definition}
\newtheorem*{defn*}{Definition}

\PassOptionsToPackage{linecolor=blue,backgroundcolor=blue!25,bordercolor=blue,textsize=scriptsize}{todonotes}
\usepackage{todonotes}

\usepackage{slashed}	 	
\usepackage{amsfonts} 		
\SetTracking{encoding={*}, shape=sc}{0}
\allowdisplaybreaks		

\makeatletter
\g@addto@macro\bfseries{\boldmath}
\makeatother

\let\originalleft\left
\let\originalright\right
\renewcommand{\left}{\mathopen{}\mathclose\bgroup\originalleft}
\renewcommand{\right}{\aftergroup\egroup\originalright}

\newcommand{\newsection}[1]{\noindent \textbf{#1}.}

\begin{document}


\preprint{Imperial/TP/21/ET/1}

\title{Exactly Marginal Deformations and their Supergravity Duals}

\author{Anthony Ashmore}
\email{ashmore@uchicago.edu}
\affiliation{Enrico Fermi Institute \& Kadanoff Center for Theoretical Physics, University of Chicago, Chicago, IL 60637, USA} 
\affiliation{Sorbonne Universit\'e, UPMC Paris 06, UMR 7589, LPTHE, 75005 Paris, France}

\author{Michela Petrini}
\email{petrini@lpthe.jussieu.fr}
\affiliation{Sorbonne Universit\'e, UPMC Paris 06, UMR 7589, LPTHE, 75005 Paris, France}

\author{Edward Tasker}
\affiliation{Department of Physics, Imperial College London, Prince Consort Road, London, SW7 2AZ, UK} 

\author{Daniel Waldram}
\email{d.waldram@imperial.ac.uk}
\affiliation{Department of Physics, Imperial College London, Prince Consort Road, London, SW7 2AZ, UK}


\begin{abstract}
	\noindent We study the space of supersymmetric AdS$_5$ solutions of type IIB supergravity corresponding to the conformal manifold of the dual $\mathcal{N}=1$ conformal field theory. We show that the background geometry naturally encodes a generalised holomorphic structure, dual to the superpotential of the field theory, with the existence of the full solution following from a continuity argument. In particular, this provides a solution to the long-standing problem of finding the gravity dual of the generic $\mathcal{N}=1$ deformations of $\mathcal{N}=4$ conformal field theory. Using this formalism, we derive a new result for the Hilbert series of the deformed field theories.
\end{abstract}

\pacs{Valid PACS appear here}


\maketitle




\newsection{Introduction} Over the last two decades, the AdS/CFT correspondence~\cite{Maldacena:1997re} has provided remarkable insights into the properties of superconformal field theories (SCFTs). For the canonical example of four-dimensional $\mathcal{N}=1$ SCFTs, it gives a dual description in terms of a type II (or eleven-dimensional) supergravity background that is a product $\AdS5\times M$ of five-dimensional anti-de Sitter space ($\AdS5$) with a five- (or six-) dimensional internal space $M$. Supersymmetry constrains the internal geometry. For example, in the case of type IIB supergravity, with only the metric and five-form flux non-trivial, $M$ is required to be a Sasaki--Einstein (SE) manifold.

Certain properties of $\mathcal{N}=1$ SCFTs depend only on holomorphic data. For example, the set of gauge-invariant operators in chiral supermultiplets forms a complex ring, determined by the superpotential of the theory $\mathcal{W}$. Classically, $\mathcal{W}$ is a holomorphic function of the chiral matter fields and in a suitable scheme it is not renormalised. For an SCFT dual to an SE geometry, the ring of (single trace) mesonic chiral operators $\tr\mathcal{O}_f$ (those built from chiral matter fields) then elegantly translates into the ring of holomorphic functions $f$ on the cone over $M$, which is, by definition, Calabi--Yau.

SCFTs often come in continuous families related by perturbations by exactly marginal operators, that is operators with zero anomalous scaling dimension even after quantum corrections. Such a family describes a ``conformal manifold'' within the space of all couplings. The canonical example is the $\mathcal{N}=1$ deformations of $\mathcal{N}=4$ $\SU{N}$ super Yang--Mills theory (SYM)~\cite{Leigh:1995ep}. In addition to deformations of the $\mathcal{N}=4$ coupling constant, there are two superpotential deformations, parametrised by 
\begin{multline}
   \label{eq:N=4marginal}
   \Delta\mathcal{W}
      = f_\beta \tr \left(\Phi^1\Phi^2\Phi^3
          +\Phi^3\Phi^2\Phi^1\right) \\
      + f_\lambda \tr \left((\Phi^1)^3
          + (\Phi^2)^3 + (\Phi^3)^3 \right),
\end{multline}
where $\Phi^a$, with $a=1,2,3$, are adjoint matter fields of $\mathcal{N}=4$ SYM written as three chiral $\mathcal{N}=1$ multiplets.

A long-standing problem has been to find the type IIB background dual to any point in the family of $\mathcal{N}=1$ deformations of $\mathcal{N}=4$. Although the duals with $f_\lambda=0$, the so-called ``beta deformations'', were found fifteen years ago by Lunin and Maldacena~\cite{Lunin:2005jy} using a beautiful solution-generating technique, to date the full dual geometry for the generic deformation is unknown. (A \emph{tour de force} perturbative calculation found the solution to second order in~\cite{Aharony:2002hx}.) The internal space always has the topology of a five-sphere, but only at the $\mathcal{N}=4$ point is the geometry simple, with only five-form flux and the round metric on $\text{S}^5$. The difficulty is that away from this point, generically all the supergravity fields must be non-trivial and all the continuous isometries of the sphere, other than the circle action generating the $U(1)_{\text{R}}$ symmetry common to all $\mathcal{N}=1$ SCFTs, are broken. 

This letter gives a solution to this problem, together with its extension to the case where $\text{S}^5$ is replaced by any SE geometry.  Details will appear in~\cite{long-paper}. Following~\cite{Ashmore:2016qvs}, we use the description of the internal space $M$ in terms of $\ExR6$ generalised geometry. The geometry of $M$ is characterised by a pair of objects $(X,K)$ that from the point of view of the five-dimensional external space transform in hyper- and vector-multiplets. Our first observation is that, quite generally, fixing the superpotential of the dual field theory fixes a certain class of hypermultiplet structures $[X]$. 

The second step is one familiar from other problems in geometry. The lack of isometries means there is little hope of finding the explicit solution $(X,K)$ that describes the generic deformation. Instead, we find an ``exceptionally deformed'' solution $(\Xed,K)$ satisfying slightly weaker, but non-trivial, conditions which translate to an $\mathcal{N}=1$ theory with the same superpotential as the deformed theory but which is not conformal. We then use continuity arguments to show that there must exist a solution $(X,K)$, in the same class $[X]=[\Xed]$, that satisfies the full set of conditions. Physically, it represents the end point of the renormalisation group flow from the non-conformal point dual to $(\Xed,K)$. This is very analogous to the analysis of Calabi--Yau metrics. There one considers a set of Kähler metrics with fixed complex structure and Kähler class, and the Calabi--Yau theorem implies there is a unique Ricci-flat metric within the class.

Since the class $[X]$ is uniquely fixed by the holomorphic data of the SCFT (the superpotential), we should be able to find the dual description of any of the holomorphic properties of the theory from $\Xed$. In particular, we derive a new result for the Hilbert series $\tilde{H}(t)$, that is the generating function for the number $n_k$ of single trace mesonic operators of  R-charge $\frac23k$ (and hence conformal dimension $k$), for generic deformations of a theory dual to any SE geometry. The general expression is given in terms of a new cohomology calculated by one of us in~\cite{ed-paper}, and can be written as
\begin{equation}
\tilde{H}(t) \coloneqq 1 + \mathcal{I}_{\text{s.t.}}(t) - [k \equiv_3 0, k>0]t^{2k} ,
\end{equation}
where $\mathcal{I}_{\text{s.t.}}(t)$ is the single-trace superconformal index and the ``Iverson bracket'' notation is defined in \eqref{eq:iverson}. For $\text{S}^5$, it is natural to write $\tilde{H}(t)=H(t^2)$, giving
\begin{equation}
   \label{eq:S5-hilbert}
   H(t) = \sum_k n_k t^k = \frac{(1+t)^3}{1-t^3} ,
\end{equation}
in agreement with the field theory expression, namely the ``reduced cyclic homology''~\cite{Eager:2012hx}, as calculated in~\cite{MR1291019}.

\vspace{2pt}
\newsection{Holomorphic structure} Let us briefly review the results of~\cite{Ashmore:2016qvs}. We consider generic type IIB solutions of the form $\AdS{5}\times M$ preserving eight supercharges using the conventions of~\cite{Ashmore:2015joa}. (All $\ExR6$ tensor expressions actually apply equally well to $\AdS5$ solutions in M-theory.) There is a warped metric
\begin{equation}
\label{eq:metric_ansatz}
  \dd s^2 = \ee^{2\Delta} \dd s^2(\AdS{5}) + \dd s^2 (M),
\end{equation}
with non-trivial axion-dilaton $\tau=C_0+\ii\ee^{-\phi}$, a doublet $i=1,2$ of three-form fluxes $F^i=\dd B^i$ and a five-form flux $F=\dd C+\frac{1}{2}\epsilon_{ij}B^i\wedge F^j\coloneqq f\vol_M$ on $M$ (with dual five-form flux $f\vol_{\AdS{5}}$ on $\AdS{5}$). We normalise the AdS metric to have unit radius $R^{\AdS{5}}_{\mu\nu}=-4 g^{\AdS{5}}_{\mu\nu}$. The bosonic symmetry of the supergravity combines diffeomorphisms and gauge transformations $\delta B^i=\dd\lambda^i$, $\delta C=\dd\rho-\frac12\epsilon_{ij}\dd\lambda^i\wedge B^j$, into a  ``generalised diffeomorphism'' group $\GDiff$. 

In the special case of $F^i=0$ and $\tau$ and $\Delta$ constant, the metric on $M$ is SE~\cite{Acharya:1998db}. The geometry is defined by a vector field $\xi$, a real one-form $\sigma$, a real two-form $\omega$ and a complex two-form $\Omega$ satisfying 
\begin{equation}
   \label{eq:SE-alg}
   \begin{aligned}
      \imath_{\xi}\sigma &=1, &
      \imath_\xi\omega=\imath_\xi\Omega &= 0, &
      \omega\wedge\omega = \tfrac{1}{2}\Omega\wedge\bar\Omega ,
   \end{aligned}
\end{equation}
and the differential conditions
\begin{equation}
   \label{eq:SE-int}
   \begin{aligned}
      \dd\sigma& = 2 \omega, & \dd\Omega=3\ii\sigma\wedge\Omega .
   \end{aligned}
\end{equation}
Mathematically $(\xi,\sigma,\omega,\Omega)$ define an $\SU2\subset\GL{5,\bbR}$ structure with singlet intrinsic torsion. 

The generic solution defines an ``exceptional Sasaki--Einstein structure'' (ExSE) on $M$~\cite{Ashmore:2016qvs}. It is encoded by a pair of generalised tensors $(X,K)$ that are combinations of conventional $\GL{5,\bbR}$ tensors, transforming under an enlarged structure group $\ExR6\supset\GL{5,\bbR}$. $K$ is a real generalised vector that defines a ``vector-multiplet structure''. A generic generalised vector $V$ is a section of~\cite{Hull:2007zu,Pacheco:2008ps}
\begin{equation}
   \label{eq:E}
   \begin{aligned}
      E &\simeq T \oplus (T^*\oplus T^*) \oplus \ext^3T^*
      \oplus (\ext^5T^* \oplus \ext^5T^*), \\
    V &= v + \lambda^i + \rho + \sigma^i \in \Gamma(E) ,
   \end{aligned}
\end{equation}
transforming in the $\rep{27}_1$ representation, where $T$ and $T^*$ are the tangent and cotangent bundles of $M$ and $v$ is a vector, $\lambda^i$ a pair of one-forms and so on. The subscript denotes the $\bbR^+$ weight, normalised such that $\det T^*$ has weight three. We will also consider sections $A$ of a bundle
\begin{equation}
   \label{eq:ad}
   \begin{aligned}
   \ad\tilde{F} &\simeq (\bbR\oplus \bbR\oplus \bbR)
      \oplus (T\otimes T^*) 
      \oplus (\ext^2T\oplus \ext^2T) \\ &\qquad
      \oplus (\ext^2T^*\oplus \ext^2T^*)
      \oplus \ext^4T T\oplus \ext^4T^* \\
   A &= a^i{}_j + r + \beta^i + b^i + \gamma + c
         \in \Gamma(\ad\tilde{F})
   \end{aligned}
\end{equation}
transforming in the $\rep{78}_0$ adjoint representation~\cite{Pacheco:2008ps,Grana:2009im,Coimbra:2011ky} and where $a^i{}_j$ is an element of the $\sln{2,\bbR}$ S-duality Lie algebra, $r$ is a $\gl{5,\bbR}$ matrix, $\beta^i$ is a doublet of bivectors, and so on. In general, we write $A\cdot{}$ for the $\ex{6(6)}\oplus\mathbb{R}$ adjoint action on any generalised tensor. The complex tensor $X$ defines a ``hypermultiplet structure'' and is a section (of the complexification) of the weighted $\rep{78}_3$ adjoint bundle
\begin{equation}
   \label{eq:ad3}
   \det T^*\otimes \ad\tilde{F} \simeq
      T^*\oplus (\ext^3T^*\oplus \ext^3T^*) \oplus \dots ,
\end{equation}
where we have used $\det T^*\otimes \ext^pT\simeq \ext^{5-p}T^*$ and the dots represent tensor bundles with forms of higher degree. Together $(X,K)$ are invariant under $\USp6\subset\ExR6$, implying the algebraic conditions, analogues of~\eqref{eq:SE-alg},
\begin{equation}
   \label{eq:ESE-alg}
   \begin{aligned}
      X\cdot K &= 0 , &
      \tr X\bar{X} &= c(K,K,K)^2 ,
   \end{aligned}
\end{equation}
where $\tr$ is the trace in the adjoint and $c(K,K,K)$ is the symmetric cubic invariant of $\ExS6$, both given in~\cite{Ashmore:2016qvs}. In addition, writing $\tr X\bar{X}=\kappa^4$ and $X=\kappa(J_1+\ii J_2)$ defines a (highest root) $\su2\subset\ex{6(6)}$ algebra $\{J_\alpha\}$ with $\alpha=1,2,3$, satisfying $[J_\alpha,J_\beta]= 2\kappa\epsilon_{\alpha\beta\gamma}J_\gamma$. 

In generalised geometry $\GDiff$ is generated infinitesimally by the generalised Lie derivative~\cite{Pacheco:2008ps,Berman:2011cg,Coimbra:2011ky}. Acting on generalised tensors it is given by the operator
\begin{equation}
   \label{eq:Dorf}
   \Dorf_V = \mathcal{L}_v - (\dd\lambda^i+\dd\rho)\cdot,
\end{equation}
where $V$ is a section of~\eqref{eq:E}, $\mathcal{L}_v$ is the usual Lie derivative and $\dd\lambda^i+\dd\rho$ acts as a section of~\eqref{eq:ad}. The Killing spinor equations for the background are equivalent to differential conditions on $(X,K)$ given by
\begin{align}
   \Dorf_K K &= 0 , & \mu_+(V) &= 0 , \label{eq:ESE1} \\
   \Dorf_KX &= 3\ii X, & \mu_3(V) &= \int_M c(K,K,V) , \label{eq:ESE2}
\end{align}
holding for all generalised vectors $V$. The objects
\begin{equation}
   \label{eq:mmap}
   \mu_\alpha(V) = -\tfrac12\epsilon_{\alpha\beta\gamma}
       \int_M \tr J_\beta \Dorf_V J_\gamma
\end{equation}
are formally a triple of moment maps for the action of $\GDiff$ on the (hyper-Kähler) infinite-dimensional space $\mathcal{Z}$ of structures $X$. Geometrically, $(X,K)$ define a generalised $\USp6$ structure with singlet generalised intrinsic torsion~\cite{Coimbra:2014uxa}. The generalised Lie derivative $\tfrac23\Dorf_K$ generates an isometry dual to the SCFT R-symmetry. 

For the special case of a SE background
\begin{equation}
   \label{eq:SE}
   \begin{aligned}
      K &= \ee^C\cdot \left( \xi - \sigma\wedge\omega \right) , \\
      X &= - \ee^{C+\frac12\ii\omega\wedge\omega} \cdot
         v^i \sigma\wedge\Omega , 
   \end{aligned}
\end{equation}
where the four-forms act by the exponentiated $\ex{6(6)}$ adjoint action, $v^i=(\tau_0,1)^i/\sqrt{\im\tau_0}$ for constant axion-dilaton $\tau_0$, and $F=\dd C=2\sigma\wedge\omega\wedge\omega$.

What part of the structure $(X,K)$ encodes the holomorphic information of the SCFT? Our claim is that it is independent of $K$ and depends only on the class
\begin{equation}
   \label{eq:classX}
   [X] = \{ \tilde{X}=g^* X : g \in \GDiff_\bbC \}
\end{equation}
of hypermultiplet structures $\tilde{X}$ related to $X$ by a complexified generalised diffeomorphism. (Strictly speaking $\GDiff_\bbC$ does not form a group and $[X]$ is actually defined as the orbit of $X$ generated by $\Dorf_V$ for all complex $V$.)

The first argument for this identification comes from considering generic non-conformal supersymmetric deformations of the SCFT. These are of either  ``superpotential'' or ``Kähler potential'' type, corresponding to the highest component of a chiral or general vector supermultiplet (see for example~\cite{Green:2010da}). In the dual gravity theory at the $\AdS5$ point, the former are dual to a 5d hypermultiplet and the latter to a 5d vector eating a hypermultiplet to become massive. Formally $X$ and $K$ can be viewed as an infinite set of hyper and vector multiplets respectively in a 5d supergravity gauged by $\GDiff$~\cite{Ashmore:2015joa,Ashmore:2016qvs}. The hypermultiplet deformations that are eaten to form a long vector are then of the form $\delta X=\Dorf_VX$ for complex $V$. Since it is only the superpotential deformations that deform the holomorphic structure of the field theory, we see we need to consider deformations of $X$ modulo those of the form $\Dorf_VX$ as in~\eqref{eq:classX}.

The second argument comes from considering the supergravity domain-wall flow equations where the $\AdS5$ metric in~\eqref{eq:metric_ansatz} is replaced by a foliation of Minkowski spaces
\begin{equation}
   \label{eq:DMmetric}
   \dd s^2 = a^2(r)\eta_{\mu\nu}\dd x^\mu\dd x^\nu + \dd r^2 .
\end{equation}
By using the analysis of~\cite{Ceresole:2001wi}, we can show that supersymmetry of the domain-wall solution implies that \eqref{eq:ESE1} hold together with the flow equations
\begin{equation}
   \label{eq:DMflow}
   \begin{aligned}
      X' &= -\tfrac{2}{3}\ii \Dorf_K X , &
      \int_M c(K,K',V) &= \mu_3(V) , 
   \end{aligned}
\end{equation}
for all $V$, where ${}'=\dd/\dd r$. Under the AdS/CFT correspondence, these describe the renormalisation group flow of a non-conformal supersymmetric $\mathcal{N}=1$ field theory, with $r$ playing the role of the energy scale. We see that $X$ flows by an \emph{imaginary} generalised diffeomorphism. Given a supersymmetric scheme, the superpotential should not flow, but this precisely implies that the holomorphic information is encoded in the class $[X]$ rather than $X$ itself. 


\vspace{2pt}
\newsection{Marginal deformations} We define an ``exceptional Sasaki'' (ExS) geometry by the slightly weaker set of conditions
\begin{equation}
   \label{eq:ES}
   \begin{aligned}
      \Dorf_K K &= 0 , & 
      \Dorf_KX &= 3\ii X, &
      \mu_+(V) &= 0 , 
   \end{aligned}
\end{equation}
where we also drop the constraint $\tr X\bar{X}=c(K,K,K)^2$. In particular, conventional Sasaski geometries are examples of ExS spaces. We now show that given a SE background~\eqref{eq:SE} and a choice function of $f$ one can construct an ExS solution. The tangent space on a SE manifold decomposes as $T_\bbC\simeq\bbC\xi\oplus T_{1,0}\oplus T_{0,1}$ with the corresponding decomposition of the exterior derivative 
\begin{equation}
   \label{eq:d-decomp}
   \dd f = \sigma\wedge\mathcal{L}_\xi f + \del f + \delb f .
\end{equation}
The action of $\delb$ on forms in $\ext^pT^*_{1,0}\otimes\ext^qT^*_{0,1}$, of charge $\ii k$ under $\mathcal{L}_\xi$, defines the ``transverse'' cohomology groups $H^{(p,q)}_{\delb}(k)$~\cite{Tievsky08,ed-paper}. If $f$ is holomorphic on the Calabi--Yau cone then $k\geq0$ and $\delb f=0$. Given such a function we define
\begin{equation}
   \label{eq:Xed}
   \Xed \coloneqq \ee^{C+\frac12\ii\omega\wedge\omega} \cdot
      \ee^{\epsilon r^ir_j+\alpha r^i}\cdot\left(
         \dd f - v^i\sigma\wedge\Omega\right) ,
\end{equation}
where $r^i=(1,0)^i$, $r_i=\epsilon_{ij}r^j$, the two-form $\alpha$ is given by  
\begin{equation}
   \label{eq:alpha}
   \alpha = - \frac{\ii}{2(k-1)}\sigma\wedge a
       - \frac{k}{k-1} f\bar{\Omega} ,
\end{equation}
where, writing $(\beta^\sharp)^m=g^{mn}\beta_n$ for a one-form $\beta$, $a=\imath_{\dd f^\sharp}\bar{\Omega}$, and $\epsilon$ satisfies
\begin{equation}
   \label{eq:epsilon}
   \delb\epsilon = - \frac{\ii}{8(k-1)} \imath_{a^\sharp}(\del a) .
\end{equation}
The RHS of~\eqref{eq:epsilon} is $\delb$-closed and, since $H^{(0,1)}(k)=0$ on a SE manifold, there is always a solution for $\epsilon$. In particular, for $\text{S}^5$ we have explicitly
\begin{equation}
\label{eq:S5epsilon}
\epsilon = \frac{\ii}{4!2^7(k-1)^2} (\omega\wedge\omega)^{mnpq}
(\del a\wedge\del a)_{mnpq}. 
\end{equation}

For general $f$, one can show that $(\Xed,K)$ satisfies the supersymmetric domain-wall conditions~\eqref{eq:ESE1}, and so characterises the dual of an $\mathcal{N}=1$ field theory. (In the language of~\cite{Tomasiello:2007zq,Gabella:2009wu} we are solving all but the $\re\Phi_+$ equation.) In particular, we claim $[\Xed]$ is the dual of a finite deformation of the original superpotential by the mesonic operator $ \Delta\mathcal{W} = \tr\mathcal{O}_f$. At linear order in $f$, one can check that $\Xed$ gives the deformations dual to $\tr\mathcal{O}_f$ derived in~\cite{Ashmore:2016oug}.

Since $\Dorf_K f=\mathcal{L}_\xi f$, the deformation has R-charge $\frac23k$ implying that if $k=3$ we have an ExS background, dual to an exactly marginal deformation. For the exactly marginal deformations of $\text{S}^5$, $f$ is a cubic function on the $\bbC^3$ cone. Working to second order, we see that $\epsilon$ deforms the axion-dilaton $\tau$, and matches the expression given in~\cite{Aharony:2002hx}. The undeformed solution is invariant under an $\SU3\subset\GDiff$ group of conventional diffeomorphisms, thus $\Xed$ define equivalent classes $[\Xed]$ under $\SU3^\bbC=\SL{3,\bbC}$ transformations of $f$, just as in the field theory~\cite{Kol:2010ub,Green:2010da}.

Since we have not satisfied the $\mu_3$ condition in~\eqref{eq:ESE2}, even in the exactly marginal case $K$ will flow under the domain-wall equations~\eqref{eq:DMflow}. Physically this corresponds to the flow of the field theory Kähler potential~\cite{Green:2010da}, and indeed $K'$ is given by a moment map quadratic in $X$ matching the field theory expression. Since the Kähler deformations are irrelevant,  physically if $k>3$ we expect the solution to flow back to the original undeformed SE background, and if $k=3$ the ExS background should flow to a unique new ExSE solution $(X,K)$ with $[X]=[\Xed]$.   

For the case of the beta deformation, we can indeed show that the solution of~\cite{Lunin:2005jy} is a $\GDiff_\bbC$ transformation of $(\Xed,K)$ that leaves $K$ invariant~\cite{long-paper}. More generally, although we cannot find the ExSE background explicitly, we can make a continuity argument that it exists. There is an important relation between moment maps and Geometric Invariant Theory (GIT) that underlies, for example, the remarkable theorems on the existence of solutions of the Hermitian Yang--Mills equations~\cite{Donaldson85,UY86} or of Kähler--Einstein metrics~\cite{KEproof}. Let $G_K\subset\GDiff$ be the one-dimensional R-symmetry group generated by $\Dorf_K$ (isomorphic to the diffeomorphism subgroup generated by $\mathcal{L}_\xi$) and $\GDiff_K$ be the centraliser of $G_K$ in $\GDiff$. The space  $\mathcal{Z}_K\subset\mathcal{Z}$ of structures $X$ satisfying the ExS conditions~\eqref{eq:ES} with fixed $K$ inherits a $\GDiff_K$-invariant Kähler metric from the hyper-Kähler metric on $\mathcal{Z}$. Furthermore, 
\begin{equation}
   \label{eq:muK}
   \mu_K(V) \coloneqq \mu_3(V)-\int_M c(K,K,V)
\end{equation}
is a moment map for the action of $\GDiff_K$ on $\mathcal{Z}_K$ such that $\mu_K=0$ gives an ExSE background. The Kempf--Ness theorem implies that there is an open subset of ``stable'' points $\mathcal{Z}_K^{\text{s}}\subset\mathcal{Z}_K$ that lie on complexified $\GDiff_K^\bbC$ orbits that intersect $\mu_K=0$ at unique solutions (up to the action of $\GDiff_K$). As we scale the function $f$ in $(\Xed,K)$ we get a continuous one-parameter family of ExS solutions, that, from~\cite{Ashmore:2016oug}, match the infinitesimal exactly marginal solutions for small $f$. Since $\mathcal{Z}_K^{\text{s}}$ is open we can expect that for a finite range of $f$ all these solutions are stable, and so can all be mapped to a ExSE solution $(X,K)$ by a $\GDiff_K^\bbC$ transformation. It is easy to show that no two $\Xed$ solutions are related by a $\GDiff_K^\bbC$ transformation, and hence each $\Xed$ solution flows to a unique ExSE solution. (The exception, as in the $\text{S}^5$ case, is when the original SE solution admits an isometry preserving $\sigma\wedge\Omega$. In this case, any two $\Xed$ solutions related by an isometry transformation define the same exactly marginal deformation~\cite{Ashmore:2016oug}, reproducing the field theory result of~\cite{Kol:2010ub,Green:2010da}.)

\vspace{2pt}
\newsection{The Hilbert series} Using our formalism we can calculate the set of single-trace mesonic operators of the deformed theories, giving new predictions for a large class of deformed $\mathcal{N}=1$ SCFTs. As a check, we compare with the known result for the $\text{S}^5$ case. The mesonic operators are chiral so can act as deformations of the superpotential, though of course these deformations are not necessarily marginal. As such we need to find deformations $\delta X$ that preserve the domain-wall conditions~\eqref{eq:ESE1}, that is $\delta\mu_+=0$. Since the superpotential is determined by the class $[X]$, if $\delta X=\Dorf_V X$ for some complex $V$, we have a trivial deformation, and so we have a cohomology
\begin{equation}
   \label{eq:mesonic-cohom}
   \text{chiral ops.}
      = \{ \delta\mu_+=0 \} \quotient \{ \delta X=\Dorf_VX \} . 
\end{equation}
We can grade the cohomology by the action of $\Dorf_K$ to identify the number of operators of a given R-charge $\frac23k$. The resulting cohomologies are non-zero for both positive and negative $k$, and count both supersymmetric deformations and supersymmetric vevs of the corresponding chiral operators. Furthermore, they include both single  mesonic operators $\tr \mathcal{O}$ and chiral operators of the form $\tr W_\alpha W^\alpha\mathcal{O}$ where $W_\alpha$ are gauge field strength superfields.   

It is straightforward to see that the cohomology depends only on the class $[X]$, reflecting the fact that the number of chiral operators depends only on the holomorphic information in the field theory. Hence we can calculate it knowing only $\Xed$. If we furthermore impose a regularity condition that $\eta\coloneqq\dd f$ is nowhere vanishing, we can then write $\Xed$ in the form 
\begin{equation}
   \label{eq:eta-Xed}
   \Xed = \ee^{c+b^i} \cdot \eta,
\end{equation}
where the complex four- and two-forms $c$ and $b^i$ satisfy
\begin{equation}
   \label{eq:complex-flux}
   \begin{aligned}
      \dd c + \tfrac12 \epsilon_{ij}b^i\wedge\dd b^j &= 0 , &
      \eta\wedge\dd b^i &= 0 . 
   \end{aligned}
\end{equation}
One can then show that solutions to~\eqref{eq:mesonic-cohom} of R-charge $r=\frac23k$ come from perturbing $b^i$ and are counted by a new graded cohomology $H_{\dd_\eta}^p(k)$ (with $p=2$) defined by the maps
\begin{equation}
   \label{eq:eta-complex}
   \begin{tikzcd}
      \Gamma(\eta\wedge\ext^{p-1}T^*M_\bbC)
      \arrow[r,"\dd"] & \Gamma(\eta\wedge\ext^p T^*M_\bbC) .
   \end{tikzcd}
\end{equation}
General properties of these groups and their expressions in terms of the transverse and Kohn--Rossi cohomologies defined by the underlying Sasaki--Einstein space will be given in~\cite{ed-paper}. Given that the R-charge of $\tr W_\alpha W^\alpha\mathcal{O}$ is two units more than that of $\tr\mathcal{O}$ and using the results of~\cite{ed-paper}, one can then derive a universal expression for the Hilbert series\footnote{In defining $\tilde{H}(t)$ we use the same power of twice the conformal dimension $t^{2k}$ that appears in the index. In examples where the R-symmetry is compact, one usually normalises by the minimal $\Uni1$ charge. These normalisations do not generally match.} of the single-trace mesonic operators
\begin{equation}
   \label{eq:gen-Hilbert}
   \tilde{H}(t) = 1 + \mathcal{I}_{\text{s.t.}}(t) - [k \equiv_3 0, k>0]t^{2k} ,
\end{equation}
where $\mathcal{I}_{\text{s.t.}}(t)$ is the single-trace superconformal index and we use Iverson bracket notation
\begin{equation}\label{eq:iverson}
   [k \equiv_3 0, k>0] = \begin{cases}
      1 & \text{if $k>0$ and $k\equiv0 \;(\operatorname{mod}3)$,} \\
      0 & \text{otherwise.}
   \end{cases} 
\end{equation}
By definition, the index is independent of the marginal deformation and is given in terms of Kohn--Rossi (or transverse) cohomology groups on the SE solution~\cite{Eager:2012hx}. Thus the Hilbert series, although in general different from the series at the undeformed point, is also independent of the particular marginal deformation, as expected since it counts short operators with protected conformal dimension, so can change only at discrete points in the moduli space. (Note that we are also only capturing operators dual to supergravity modes, so miss extra operators dual to wrapped string states, as for example in~\cite{Lunin:2005jy}, that are expected to appear when there is an algebraic relation between the marginal couplings.) For $\text{S}^5$ the expression~\eqref{eq:gen-Hilbert} reduces to~\eqref{eq:S5-hilbert}, and applies to generic deformations but notably not the beta deformation, since in the latter case $\eta$ vanishes on the three lines in the $\bbC^3$ cone: $z^1=z^2=0$, $z^2=z^3=0$ and $z^3=z^1=0$. 

As discussed in~\cite{Berenstein:2000ux,Eager:2012hx}, the counting of mesonic operators is given by the dimensions of the reduced cyclic homology group $\HCr_0(\mathcal{A})$ of a non-commutative ``Calabi--Yau algebra'' $\mathcal{A}$ defined by the SCFT with its superpotential~\cite{Ginzburg:2006fu}. For $\text{S}^5$ the corresponding algebras are of Sklyanin-type. The reduced cyclic homology has been calculated in the mathematics literature~\cite{MR1291019} and is in agreement with~\eqref{eq:S5-hilbert}. We have also checked the first few terms of the general expression~\eqref{eq:gen-Hilbert} for some other simple examples (such as the conifold $T^{1,1}$). More generally, it is a prediction for the set of mesonic operators for deformations of any theory away from the SE point, given non-vanishing $\eta$.

\vspace{2pt}
\newsection{Final comments} The essential ingredient of our construction is the generalised structure $X$ that encodes the holomorphic information of the dual field theory. This structure is present for any $d=3,4$ CFT with at least four supercharges, so in particular characterises the Pilch--Warner solution in type IIB, and AdS$_4$ and AdS$_5$ backgrounds in M-theory. This perspective might be especially useful for analysing M5-branes wrapped on Calabi--Yau threefolds~\cite{Maldacena:2000mw}, and for describing $F$- or $a$-maximisation away from the Sasaki--Einstein limit.


\vspace{2pt}
\newsection{Acknowledgements} AA is supported by the European Union's Horizon 2020 research and innovation programme under the Marie Sk\l{}odowska-Curie grant agreement No.~838776. ET is supported by an STFC PhD studentship. DW is supported in part by the STFC Consolidated Grant ST/T000791/1 and the EPSRC New Horizons Grant EP/V049089/1. We acknowledge the Mainz Institute for Theoretical Physics (MITP) of the Cluster of Excellence PRISMA+ (Project ID 39083149) for hospitality and support during part of this work.

\vspace{2pt}
\newsection{Note added}
Edward Tasker passed away in January 2020. He made major contributions to this work during his PhD studies at Imperial College London. 

\smallskip

\noindent Ed was a much-loved colleague and friend, and a gifted physicist and mathematician with a seemingly endless supply of puns and a knack for solving problems in unexpected ways. We miss him greatly. We hope this work will add to his memory. (AA, MP and DW.)



\begin{thebibliography}{30}%
\makeatletter
\providecommand \@ifxundefined [1]{%
 \@ifx{#1\undefined}
}%
\providecommand \@ifnum [1]{%
 \ifnum #1\expandafter \@firstoftwo
 \else \expandafter \@secondoftwo
 \fi
}%
\providecommand \@ifx [1]{%
 \ifx #1\expandafter \@firstoftwo
 \else \expandafter \@secondoftwo
 \fi
}%
\providecommand \natexlab [1]{#1}%
\providecommand \enquote  [1]{``#1''}%
\providecommand \bibnamefont  [1]{#1}%
\providecommand \bibfnamefont [1]{#1}%
\providecommand \citenamefont [1]{#1}%
\providecommand \href@noop [0]{\@secondoftwo}%
\providecommand \href [0]{\begingroup \@sanitize@url \@href}%
\providecommand \@href[1]{\@@startlink{#1}\@@href}%
\providecommand \@@href[1]{\endgroup#1\@@endlink}%
\providecommand \@sanitize@url [0]{\catcode `\\12\catcode `\$12\catcode
  `\&12\catcode `\#12\catcode `\^12\catcode `\_12\catcode `\%12\relax}%
\providecommand \@@startlink[1]{}%
\providecommand \@@endlink[0]{}%
\providecommand \url  [0]{\begingroup\@sanitize@url \@url }%
\providecommand \@url [1]{\endgroup\@href {#1}{\urlprefix }}%
\providecommand \urlprefix  [0]{URL }%
\providecommand \Eprint [0]{\href }%
\providecommand \doibase [0]{https://doi.org/}%
\providecommand \selectlanguage [0]{\@gobble}%
\providecommand \bibinfo  [0]{\@secondoftwo}%
\providecommand \bibfield  [0]{\@secondoftwo}%
\providecommand \translation [1]{[#1]}%
\providecommand \BibitemOpen [0]{}%
\providecommand \bibitemStop [0]{}%
\providecommand \bibitemNoStop [0]{.\EOS\space}%
\providecommand \EOS [0]{\spacefactor3000\relax}%
\providecommand \BibitemShut  [1]{\csname bibitem#1\endcsname}%
\let\auto@bib@innerbib\@empty
\bibitem [{\citenamefont {Maldacena}(1999)}]{Maldacena:1997re}%
  \BibitemOpen
  \bibfield  {author} {\bibinfo {author} {\bibfnamefont {J.~M.}\ \bibnamefont
  {Maldacena}},\ }\bibfield  {title} {\bibinfo {title} {{The Large $N$ Limit of
  Superconformal Field Theories and Supergravity}},\ }\href
  {https://doi.org/10.1023/A:1026654312961, 10.4310/ATMP.1998.v2.n2.a1}
  {\bibfield  {journal} {\bibinfo  {journal} {Int. J. Theor. Phys.}\ }\textbf
  {\bibinfo {volume} {38}},\ \bibinfo {pages} {1113} (\bibinfo {year}
  {1999})},\ \bibinfo {note} {[Adv. Theor. Math. Phys.2,231(1998)]},\ \Eprint
  {https://arxiv.org/abs/hep-th/9711200} {arXiv:hep-th/9711200 [hep-th]}
  \BibitemShut {NoStop}%
\bibitem [{\citenamefont {Leigh}\ and\ \citenamefont
  {Strassler}(1995)}]{Leigh:1995ep}%
  \BibitemOpen
  \bibfield  {author} {\bibinfo {author} {\bibfnamefont {R.~G.}\ \bibnamefont
  {Leigh}}\ and\ \bibinfo {author} {\bibfnamefont {M.~J.}\ \bibnamefont
  {Strassler}},\ }\bibfield  {title} {\bibinfo {title} {{Exactly Marginal
  Operators and Duality in Four-Dimensional ${\mathcal{N}}\!=1$ Supersymmetric
  Gauge Theory}},\ }\href {https://doi.org/10.1016/0550-3213(95)00261-P}
  {\bibfield  {journal} {\bibinfo  {journal} {Nucl. Phys.}\ }\textbf {\bibinfo
  {volume} {B447}},\ \bibinfo {pages} {95} (\bibinfo {year} {1995})},\ \Eprint
  {https://arxiv.org/abs/hep-th/9503121} {arXiv:hep-th/9503121 [hep-th]}
  \BibitemShut {NoStop}%
\bibitem [{\citenamefont {Lunin}\ and\ \citenamefont
  {Maldacena}(2005)}]{Lunin:2005jy}%
  \BibitemOpen
  \bibfield  {author} {\bibinfo {author} {\bibfnamefont {O.}~\bibnamefont
  {Lunin}}\ and\ \bibinfo {author} {\bibfnamefont {J.~M.}\ \bibnamefont
  {Maldacena}},\ }\bibfield  {title} {\bibinfo {title} {{Deforming Field
  Theories with U(1) $\times$ U(1) Global Symmetry and Their Gravity Duals}},\
  }\href {https://doi.org/10.1088/1126-6708/2005/05/033} {\bibfield  {journal}
  {\bibinfo  {journal} {JHEP}\ }\textbf {\bibinfo {volume} {05}},\ \bibinfo
  {pages} {033}},\ \Eprint {https://arxiv.org/abs/hep-th/0502086}
  {arXiv:hep-th/0502086 [hep-th]} \BibitemShut {NoStop}%
\bibitem [{\citenamefont {Aharony}\ \emph {et~al.}(2002)\citenamefont
  {Aharony}, \citenamefont {Kol},\ and\ \citenamefont
  {Yankielowicz}}]{Aharony:2002hx}%
  \BibitemOpen
  \bibfield  {author} {\bibinfo {author} {\bibfnamefont {O.}~\bibnamefont
  {Aharony}}, \bibinfo {author} {\bibfnamefont {B.}~\bibnamefont {Kol}},\ and\
  \bibinfo {author} {\bibfnamefont {S.}~\bibnamefont {Yankielowicz}},\
  }\bibfield  {title} {\bibinfo {title} {{On Exactly Marginal Deformations of
  ${\mathcal{N}}\!=4$ Sym and Type IIB Supergravity on $\mathrm{AdS}_5$
  $\times$ $ S^5$}},\ }\href {https://doi.org/10.1088/1126-6708/2002/06/039}
  {\bibfield  {journal} {\bibinfo  {journal} {JHEP}\ }\textbf {\bibinfo
  {volume} {06}},\ \bibinfo {pages} {039}},\ \Eprint
  {https://arxiv.org/abs/hep-th/0205090} {arXiv:hep-th/0205090 [hep-th]}
  \BibitemShut {NoStop}%
\bibitem [{\citenamefont {Ashmore}\ \emph {et~al.}()\citenamefont {Ashmore},
  \citenamefont {Petrini}, \citenamefont {Tasker},\ and\ \citenamefont
  {Waldram}}]{long-paper}%
  \BibitemOpen
  \bibfield  {author} {\bibinfo {author} {\bibfnamefont {A.}~\bibnamefont
  {Ashmore}}, \bibinfo {author} {\bibfnamefont {M.}~\bibnamefont {Petrini}},
  \bibinfo {author} {\bibfnamefont {E.}~\bibnamefont {Tasker}},\ and\ \bibinfo
  {author} {\bibfnamefont {D.}~\bibnamefont {Waldram}},\ }\bibinfo {note} {to
  appear}\BibitemShut {NoStop}%
\bibitem [{\citenamefont {Ashmore}\ \emph {et~al.}(2016)\citenamefont
  {Ashmore}, \citenamefont {Petrini},\ and\ \citenamefont
  {Waldram}}]{Ashmore:2016qvs}%
  \BibitemOpen
  \bibfield  {author} {\bibinfo {author} {\bibfnamefont {A.}~\bibnamefont
  {Ashmore}}, \bibinfo {author} {\bibfnamefont {M.}~\bibnamefont {Petrini}},\
  and\ \bibinfo {author} {\bibfnamefont {D.}~\bibnamefont {Waldram}},\
  }\bibfield  {title} {\bibinfo {title} {{The Exceptional Generalised Geometry
  of Supersymmetric AdS Flux Backgrounds}},\ }\href
  {https://doi.org/10.1007/JHEP12(2016)146} {\bibfield  {journal} {\bibinfo
  {journal} {JHEP}\ }\textbf {\bibinfo {volume} {12}},\ \bibinfo {pages}
  {146}},\ \Eprint {https://arxiv.org/abs/1602.02158} {arXiv:1602.02158
  [hep-th]} \BibitemShut {NoStop}%
\bibitem [{\citenamefont {Tasker}()}]{ed-paper}%
  \BibitemOpen
  \bibfield  {author} {\bibinfo {author} {\bibfnamefont {E.}~\bibnamefont
  {Tasker}},\ }\bibfield  {title} {\bibinfo {title} {From $\beta$ to $\eta$: a
  new cohomology for deformed {S}asaki--{E}instein},\ }\bibinfo {note} {to
  appear}\BibitemShut {NoStop}%
\bibitem [{\citenamefont {Eager}\ \emph {et~al.}(2014)\citenamefont {Eager},
  \citenamefont {Schmude},\ and\ \citenamefont {Tachikawa}}]{Eager:2012hx}%
  \BibitemOpen
  \bibfield  {author} {\bibinfo {author} {\bibfnamefont {R.}~\bibnamefont
  {Eager}}, \bibinfo {author} {\bibfnamefont {J.}~\bibnamefont {Schmude}},\
  and\ \bibinfo {author} {\bibfnamefont {Y.}~\bibnamefont {Tachikawa}},\
  }\bibfield  {title} {\bibinfo {title} {{Superconformal Indices,
  Sasaki-Einstein Manifolds, and Cyclic Homologies}},\ }\href
  {https://doi.org/10.4310/ATMP.2014.v18.n1.a3} {\bibfield  {journal} {\bibinfo
   {journal} {Adv. Theor. Math. Phys.}\ }\textbf {\bibinfo {volume} {18}},\
  \bibinfo {pages} {129} (\bibinfo {year} {2014})},\ \Eprint
  {https://arxiv.org/abs/1207.0573} {arXiv:1207.0573 [hep-th]} \BibitemShut
  {NoStop}%
\bibitem [{\citenamefont {Van~den Bergh}(1994)}]{MR1291019}%
  \BibitemOpen
  \bibfield  {author} {\bibinfo {author} {\bibfnamefont {M.}~\bibnamefont
  {Van~den Bergh}},\ }\bibfield  {title} {\bibinfo {title} {Noncommutative
  homology of some three-dimensional quantum spaces},\ }in\ \href
  {https://doi.org/10.1007/BF00960862} {\emph {\bibinfo {booktitle}
  {Proceedings of {C}onference on {A}lgebraic {G}eometry and {R}ing {T}heory in
  honor of {M}ichael {A}rtin, {P}art {III} ({A}ntwerp, 1992)}}},\ Vol.~\bibinfo
  {volume} {8}\ (\bibinfo {year} {1994})\ pp.\ \bibinfo {pages}
  {213--230}\BibitemShut {NoStop}%
\bibitem [{\citenamefont {Ashmore}\ and\ \citenamefont
  {Waldram}(2017)}]{Ashmore:2015joa}%
  \BibitemOpen
  \bibfield  {author} {\bibinfo {author} {\bibfnamefont {A.}~\bibnamefont
  {Ashmore}}\ and\ \bibinfo {author} {\bibfnamefont {D.}~\bibnamefont
  {Waldram}},\ }\bibfield  {title} {\bibinfo {title} {{Exceptional Calabi-Yau
  spaces: the geometry of $\mathcal{N}=2$ backgrounds with flux}},\ }\href
  {https://doi.org/10.1002/prop.201600109} {\bibfield  {journal} {\bibinfo
  {journal} {Fortsch. Phys.}\ }\textbf {\bibinfo {volume} {65}},\ \bibinfo
  {pages} {1600109} (\bibinfo {year} {2017})},\ \Eprint
  {https://arxiv.org/abs/1510.00022} {arXiv:1510.00022 [hep-th]} \BibitemShut
  {NoStop}%
\bibitem [{\citenamefont {Acharya}\ \emph {et~al.}(1999)\citenamefont
  {Acharya}, \citenamefont {Figueroa-O'Farrill}, \citenamefont {Hull},\ and\
  \citenamefont {Spence}}]{Acharya:1998db}%
  \BibitemOpen
  \bibfield  {author} {\bibinfo {author} {\bibfnamefont {B.~S.}\ \bibnamefont
  {Acharya}}, \bibinfo {author} {\bibfnamefont {J.~M.}\ \bibnamefont
  {Figueroa-O'Farrill}}, \bibinfo {author} {\bibfnamefont {C.~M.}\ \bibnamefont
  {Hull}},\ and\ \bibinfo {author} {\bibfnamefont {B.~J.}\ \bibnamefont
  {Spence}},\ }\bibfield  {title} {\bibinfo {title} {{Branes at Conical
  Singularities and Holography}},\ }\href
  {https://doi.org/10.4310/ATMP.1998.v2.n6.a2} {\bibfield  {journal} {\bibinfo
  {journal} {Adv. Theor. Math. Phys.}\ }\textbf {\bibinfo {volume} {2}},\
  \bibinfo {pages} {1249} (\bibinfo {year} {1999})},\ \Eprint
  {https://arxiv.org/abs/hep-th/9808014} {arXiv:hep-th/9808014 [hep-th]}
  \BibitemShut {NoStop}%
\bibitem [{\citenamefont {Hull}(2007)}]{Hull:2007zu}%
  \BibitemOpen
  \bibfield  {author} {\bibinfo {author} {\bibfnamefont {C.~M.}\ \bibnamefont
  {Hull}},\ }\bibfield  {title} {\bibinfo {title} {{Generalised Geometry for
  M-theory}},\ }\href {https://doi.org/10.1088/1126-6708/2007/07/079}
  {\bibfield  {journal} {\bibinfo  {journal} {JHEP}\ }\textbf {\bibinfo
  {volume} {07}},\ \bibinfo {pages} {079}},\ \Eprint
  {https://arxiv.org/abs/hep-th/0701203} {arXiv:hep-th/0701203 [hep-th]}
  \BibitemShut {NoStop}%
\bibitem [{\citenamefont {Pires~Pacheco}\ and\ \citenamefont
  {Waldram}(2008)}]{Pacheco:2008ps}%
  \BibitemOpen
  \bibfield  {author} {\bibinfo {author} {\bibfnamefont {P.}~\bibnamefont
  {Pires~Pacheco}}\ and\ \bibinfo {author} {\bibfnamefont {D.}~\bibnamefont
  {Waldram}},\ }\bibfield  {title} {\bibinfo {title} {{M-theory, Exceptional
  Generalised Geometry and Superpotentials}},\ }\href
  {https://doi.org/10.1088/1126-6708/2008/09/123} {\bibfield  {journal}
  {\bibinfo  {journal} {JHEP}\ }\textbf {\bibinfo {volume} {09}},\ \bibinfo
  {pages} {123}},\ \Eprint {https://arxiv.org/abs/0804.1362} {arXiv:0804.1362
  [hep-th]} \BibitemShut {NoStop}%
\bibitem [{\citenamefont {Gra\~na}\ \emph {et~al.}(2009)\citenamefont
  {Gra\~na}, \citenamefont {Louis}, \citenamefont {Sim},\ and\ \citenamefont
  {Waldram}}]{Grana:2009im}%
  \BibitemOpen
  \bibfield  {author} {\bibinfo {author} {\bibfnamefont {M.}~\bibnamefont
  {Gra\~na}}, \bibinfo {author} {\bibfnamefont {J.}~\bibnamefont {Louis}},
  \bibinfo {author} {\bibfnamefont {A.}~\bibnamefont {Sim}},\ and\ \bibinfo
  {author} {\bibfnamefont {D.}~\bibnamefont {Waldram}},\ }\bibfield  {title}
  {\bibinfo {title} {{E7(7) Formulation of ${\mathcal{N}}\!=2$ Backgrounds}},\
  }\href {https://doi.org/10.1088/1126-6708/2009/07/104} {\bibfield  {journal}
  {\bibinfo  {journal} {JHEP}\ }\textbf {\bibinfo {volume} {07}},\ \bibinfo
  {pages} {104}},\ \Eprint {https://arxiv.org/abs/0904.2333} {arXiv:0904.2333
  [hep-th]} \BibitemShut {NoStop}%
\bibitem [{\citenamefont {Coimbra}\ \emph {et~al.}(2014)\citenamefont
  {Coimbra}, \citenamefont {Strickland-Constable},\ and\ \citenamefont
  {Waldram}}]{Coimbra:2011ky}%
  \BibitemOpen
  \bibfield  {author} {\bibinfo {author} {\bibfnamefont {A.}~\bibnamefont
  {Coimbra}}, \bibinfo {author} {\bibfnamefont {C.}~\bibnamefont
  {Strickland-Constable}},\ and\ \bibinfo {author} {\bibfnamefont
  {D.}~\bibnamefont {Waldram}},\ }\bibfield  {title} {\bibinfo {title}
  {{$E_{d(d)} \times \mathbb{R}^+$ generalised geometry, connections and M
  theory}},\ }\href {https://doi.org/10.1007/JHEP02(2014)054} {\bibfield
  {journal} {\bibinfo  {journal} {JHEP}\ }\textbf {\bibinfo {volume} {02}},\
  \bibinfo {pages} {054}},\ \Eprint {https://arxiv.org/abs/1112.3989}
  {arXiv:1112.3989 [hep-th]} \BibitemShut {NoStop}%
\bibitem [{\citenamefont {Berman}\ \emph {et~al.}(2012)\citenamefont {Berman},
  \citenamefont {Godazgar}, \citenamefont {Godazgar},\ and\ \citenamefont
  {Perry}}]{Berman:2011cg}%
  \BibitemOpen
  \bibfield  {author} {\bibinfo {author} {\bibfnamefont {D.~S.}\ \bibnamefont
  {Berman}}, \bibinfo {author} {\bibfnamefont {H.}~\bibnamefont {Godazgar}},
  \bibinfo {author} {\bibfnamefont {M.}~\bibnamefont {Godazgar}},\ and\
  \bibinfo {author} {\bibfnamefont {M.~J.}\ \bibnamefont {Perry}},\ }\bibfield
  {title} {\bibinfo {title} {{The Local Symmetries of M-theory and Their
  Formulation in Generalised Geometry}},\ }\href
  {https://doi.org/10.1007/JHEP01(2012)012} {\bibfield  {journal} {\bibinfo
  {journal} {JHEP}\ }\textbf {\bibinfo {volume} {01}},\ \bibinfo {pages}
  {012}},\ \Eprint {https://arxiv.org/abs/1110.3930} {arXiv:1110.3930 [hep-th]}
  \BibitemShut {NoStop}%
\bibitem [{\citenamefont {Coimbra}\ \emph {et~al.}(2016)\citenamefont
  {Coimbra}, \citenamefont {Strickland-Constable},\ and\ \citenamefont
  {Waldram}}]{Coimbra:2014uxa}%
  \BibitemOpen
  \bibfield  {author} {\bibinfo {author} {\bibfnamefont {A.}~\bibnamefont
  {Coimbra}}, \bibinfo {author} {\bibfnamefont {C.}~\bibnamefont
  {Strickland-Constable}},\ and\ \bibinfo {author} {\bibfnamefont
  {D.}~\bibnamefont {Waldram}},\ }\bibfield  {title} {\bibinfo {title}
  {{Supersymmetric Backgrounds and Generalised Special Holonomy}},\ }\href
  {https://doi.org/10.1088/0264-9381/33/12/125026} {\bibfield  {journal}
  {\bibinfo  {journal} {Class. Quant. Grav.}\ }\textbf {\bibinfo {volume}
  {33}},\ \bibinfo {pages} {125026} (\bibinfo {year} {2016})},\ \Eprint
  {https://arxiv.org/abs/1411.5721} {arXiv:1411.5721 [hep-th]} \BibitemShut
  {NoStop}%
\bibitem [{\citenamefont {Green}\ \emph {et~al.}(2010)\citenamefont {Green},
  \citenamefont {Komargodski}, \citenamefont {Seiberg}, \citenamefont
  {Tachikawa},\ and\ \citenamefont {Wecht}}]{Green:2010da}%
  \BibitemOpen
  \bibfield  {author} {\bibinfo {author} {\bibfnamefont {D.}~\bibnamefont
  {Green}}, \bibinfo {author} {\bibfnamefont {Z.}~\bibnamefont {Komargodski}},
  \bibinfo {author} {\bibfnamefont {N.}~\bibnamefont {Seiberg}}, \bibinfo
  {author} {\bibfnamefont {Y.}~\bibnamefont {Tachikawa}},\ and\ \bibinfo
  {author} {\bibfnamefont {B.}~\bibnamefont {Wecht}},\ }\bibfield  {title}
  {\bibinfo {title} {{Exactly Marginal Deformations and Global Symmetries}},\
  }\href {https://doi.org/10.1007/JHEP06(2010)106} {\bibfield  {journal}
  {\bibinfo  {journal} {JHEP}\ }\textbf {\bibinfo {volume} {06}},\ \bibinfo
  {pages} {106}},\ \Eprint {https://arxiv.org/abs/1005.3546} {arXiv:1005.3546
  [hep-th]} \BibitemShut {NoStop}%
\bibitem [{\citenamefont {Ceresole}\ \emph {et~al.}(2001)\citenamefont
  {Ceresole}, \citenamefont {Dall'Agata}, \citenamefont {Kallosh},\ and\
  \citenamefont {Van~Proeyen}}]{Ceresole:2001wi}%
  \BibitemOpen
  \bibfield  {author} {\bibinfo {author} {\bibfnamefont {A.}~\bibnamefont
  {Ceresole}}, \bibinfo {author} {\bibfnamefont {G.}~\bibnamefont
  {Dall'Agata}}, \bibinfo {author} {\bibfnamefont {R.}~\bibnamefont
  {Kallosh}},\ and\ \bibinfo {author} {\bibfnamefont {A.}~\bibnamefont
  {Van~Proeyen}},\ }\bibfield  {title} {\bibinfo {title} {{Hypermultiplets,
  Domain Walls and Supersymmetric Attractors}},\ }\href
  {https://doi.org/10.1103/PhysRevD.64.104006} {\bibfield  {journal} {\bibinfo
  {journal} {Phys. Rev.}\ }\textbf {\bibinfo {volume} {D64}},\ \bibinfo {pages}
  {104006} (\bibinfo {year} {2001})},\ \Eprint
  {https://arxiv.org/abs/hep-th/0104056} {arXiv:hep-th/0104056 [hep-th]}
  \BibitemShut {NoStop}%
\bibitem [{\citenamefont {Tievsky}(2008)}]{Tievsky08}%
  \BibitemOpen
  \bibfield  {author} {\bibinfo {author} {\bibfnamefont {A.}~\bibnamefont
  {Tievsky}},\ }\emph {\bibinfo {title} {Analogues of {K}\"{a}hler geometry on
  {S}asakian manifolds}},\ \href {http://hdl.handle.net/1721.1/45349} {Ph.D.
  thesis},\ \bibinfo  {school} {Massachusetts Institute of Technology}
  (\bibinfo {year} {2008})\BibitemShut {NoStop}%
\bibitem [{\citenamefont {Tomasiello}(2008)}]{Tomasiello:2007zq}%
  \BibitemOpen
  \bibfield  {author} {\bibinfo {author} {\bibfnamefont {A.}~\bibnamefont
  {Tomasiello}},\ }\bibfield  {title} {\bibinfo {title} {{Reformulating
  Supersymmetry with a Generalized Dolbeault Operator}},\ }\href
  {https://doi.org/10.1088/1126-6708/2008/02/010} {\bibfield  {journal}
  {\bibinfo  {journal} {JHEP}\ }\textbf {\bibinfo {volume} {02}},\ \bibinfo
  {pages} {010}},\ \Eprint {https://arxiv.org/abs/0704.2613} {arXiv:0704.2613
  [hep-th]} \BibitemShut {NoStop}%
\bibitem [{\citenamefont {Gabella}\ \emph {et~al.}(2010)\citenamefont
  {Gabella}, \citenamefont {Gauntlett}, \citenamefont {Palti}, \citenamefont
  {Sparks},\ and\ \citenamefont {Waldram}}]{Gabella:2009wu}%
  \BibitemOpen
  \bibfield  {author} {\bibinfo {author} {\bibfnamefont {M.}~\bibnamefont
  {Gabella}}, \bibinfo {author} {\bibfnamefont {J.~P.}\ \bibnamefont
  {Gauntlett}}, \bibinfo {author} {\bibfnamefont {E.}~\bibnamefont {Palti}},
  \bibinfo {author} {\bibfnamefont {J.}~\bibnamefont {Sparks}},\ and\ \bibinfo
  {author} {\bibfnamefont {D.}~\bibnamefont {Waldram}},\ }\bibfield  {title}
  {\bibinfo {title} {{$\mathrm{AdS}_5$ Solutions of Type IIB Supergravity and
  Generalized Complex Geometry}},\ }\href
  {https://doi.org/10.1007/s00220-010-1083-y} {\bibfield  {journal} {\bibinfo
  {journal} {Commun. Math. Phys.}\ }\textbf {\bibinfo {volume} {299}},\
  \bibinfo {pages} {365} (\bibinfo {year} {2010})},\ \Eprint
  {https://arxiv.org/abs/0906.4109} {arXiv:0906.4109 [hep-th]} \BibitemShut
  {NoStop}%
\bibitem [{\citenamefont {Ashmore}\ \emph {et~al.}(2017)\citenamefont
  {Ashmore}, \citenamefont {Gabella}, \citenamefont {Graña}, \citenamefont
  {Petrini},\ and\ \citenamefont {Waldram}}]{Ashmore:2016oug}%
  \BibitemOpen
  \bibfield  {author} {\bibinfo {author} {\bibfnamefont {A.}~\bibnamefont
  {Ashmore}}, \bibinfo {author} {\bibfnamefont {M.}~\bibnamefont {Gabella}},
  \bibinfo {author} {\bibfnamefont {M.}~\bibnamefont {Graña}}, \bibinfo
  {author} {\bibfnamefont {M.}~\bibnamefont {Petrini}},\ and\ \bibinfo {author}
  {\bibfnamefont {D.}~\bibnamefont {Waldram}},\ }\bibfield  {title} {\bibinfo
  {title} {{Exactly marginal deformations from exceptional generalised
  geometry}},\ }\href {https://doi.org/10.1007/JHEP01(2017)124} {\bibfield
  {journal} {\bibinfo  {journal} {JHEP}\ }\textbf {\bibinfo {volume} {01}},\
  \bibinfo {pages} {124}},\ \Eprint {https://arxiv.org/abs/1605.05730}
  {arXiv:1605.05730 [hep-th]} \BibitemShut {NoStop}%
\bibitem [{\citenamefont {Kol}(2010)}]{Kol:2010ub}%
  \BibitemOpen
  \bibfield  {author} {\bibinfo {author} {\bibfnamefont {B.}~\bibnamefont
  {Kol}},\ }\bibfield  {title} {\bibinfo {title} {{On Conformal Deformations
  II}},\ }\Eprint {https://arxiv.org/abs/1005.4408} {arXiv:1005.4408 [hep-th]}
  (\bibinfo {year} {2010})\BibitemShut {NoStop}%
\bibitem [{\citenamefont {Donaldson}(1985)}]{Donaldson85}%
  \BibitemOpen
  \bibfield  {author} {\bibinfo {author} {\bibfnamefont {S.}~\bibnamefont
  {Donaldson}},\ }\bibfield  {title} {\bibinfo {title} {Anti self-dual
  {Y}ang--{M}ills connections over complex algebraic surfaces and stable vector
  bundles},\ }\href {https://doi.org/10.1112/plms/s3-50.1.1} {\bibfield
  {journal} {\bibinfo  {journal} {Proc. Lond. Math. Soc.}\ }\textbf {\bibinfo
  {volume} {50}},\ \bibinfo {pages} {1} (\bibinfo {year} {1985})}\BibitemShut
  {NoStop}%
\bibitem [{\citenamefont {Uhlenbeck}\ and\ \citenamefont {Yau}(1986)}]{UY86}%
  \BibitemOpen
  \bibfield  {author} {\bibinfo {author} {\bibfnamefont {K.}~\bibnamefont
  {Uhlenbeck}}\ and\ \bibinfo {author} {\bibfnamefont {S.-T.}\ \bibnamefont
  {Yau}},\ }\bibfield  {title} {\bibinfo {title} {On the existence of
  {H}ermitian-{Y}ang--{M}ills connections in stable vector bundles},\ }\href
  {https://doi.org/10.1002/cpa.3160390714} {\bibfield  {journal} {\bibinfo
  {journal} {Comm. Pure Appl. Math.}\ }\textbf {\bibinfo {volume} {39}},\
  \bibinfo {pages} {257} (\bibinfo {year} {1986})}\BibitemShut {NoStop}%
\bibitem [{\citenamefont {{Chen}}\ \emph {et~al.}(2015)\citenamefont {{Chen}},
  \citenamefont {{Donaldson}},\ and\ \citenamefont {{Sun}}}]{KEproof}%
  \BibitemOpen
  \bibfield  {author} {\bibinfo {author} {\bibfnamefont {X.}~\bibnamefont
  {{Chen}}}, \bibinfo {author} {\bibfnamefont {S.}~\bibnamefont
  {{Donaldson}}},\ and\ \bibinfo {author} {\bibfnamefont {S.}~\bibnamefont
  {{Sun}}},\ }\bibfield  {title} {\bibinfo {title} {{K\"ahler-Einstein metrics
  on Fano manifolds. I--III}},\ }\href@noop {} {\bibfield  {journal} {\bibinfo
  {journal} {{J. Am. Math. Soc.}}\ }\textbf {\bibinfo {volume} {28}},\ \bibinfo
  {pages} {183} (\bibinfo {year} {2015})}\BibitemShut {NoStop}%
\bibitem [{\citenamefont {Berenstein}\ \emph {et~al.}(2000)\citenamefont
  {Berenstein}, \citenamefont {Jejjala},\ and\ \citenamefont
  {Leigh}}]{Berenstein:2000ux}%
  \BibitemOpen
  \bibfield  {author} {\bibinfo {author} {\bibfnamefont {D.}~\bibnamefont
  {Berenstein}}, \bibinfo {author} {\bibfnamefont {V.}~\bibnamefont
  {Jejjala}},\ and\ \bibinfo {author} {\bibfnamefont {R.~G.}\ \bibnamefont
  {Leigh}},\ }\bibfield  {title} {\bibinfo {title} {{Marginal and Relevant
  Deformations of ${\mathcal{N}}\!=4$ Field Theories and Noncommutative Moduli
  Spaces of Vacua}},\ }\href {https://doi.org/10.1016/S0550-3213(00)00394-1}
  {\bibfield  {journal} {\bibinfo  {journal} {Nucl. Phys.}\ }\textbf {\bibinfo
  {volume} {B589}},\ \bibinfo {pages} {196} (\bibinfo {year} {2000})},\ \Eprint
  {https://arxiv.org/abs/hep-th/0005087} {arXiv:hep-th/0005087 [hep-th]}
  \BibitemShut {NoStop}%
\bibitem [{\citenamefont {Ginzburg}(2006)}]{Ginzburg:2006fu}%
  \BibitemOpen
  \bibfield  {author} {\bibinfo {author} {\bibfnamefont {V.}~\bibnamefont
  {Ginzburg}},\ }\bibfield  {title} {\bibinfo {title} {{Calabi-Yau Algebras}},\
  }\Eprint {https://arxiv.org/abs/math/0612139} {arXiv:math/0612139 [math.AG]}
  (\bibinfo {year} {2006})\BibitemShut {NoStop}%
\bibitem [{\citenamefont {Maldacena}\ and\ \citenamefont
  {Nunez}(2001)}]{Maldacena:2000mw}%
  \BibitemOpen
  \bibfield  {author} {\bibinfo {author} {\bibfnamefont {J.~M.}\ \bibnamefont
  {Maldacena}}\ and\ \bibinfo {author} {\bibfnamefont {C.}~\bibnamefont
  {Nunez}},\ }\bibfield  {title} {\bibinfo {title} {{Supergravity description
  of field theories on curved manifolds and a no go theorem}},\ }\href
  {https://doi.org/10.1142/S0217751X01003937} {\bibfield  {journal} {\bibinfo
  {journal} {Int. J. Mod. Phys. A}\ }\textbf {\bibinfo {volume} {16}},\
  \bibinfo {pages} {822} (\bibinfo {year} {2001})},\ \Eprint
  {https://arxiv.org/abs/hep-th/0007018} {arXiv:hep-th/0007018} \BibitemShut
  {NoStop}%
\end{thebibliography}
%


\end{document}